\documentclass[twocolumn,prl,aps,amsmath,showpacs,amssymb,floatfix]{revtex4-1}
\usepackage{bbm}
\usepackage{bm}
\usepackage{graphicx}
\usepackage{color}
\usepackage{amsmath,amssymb}
\usepackage{array}
\usepackage{hyperref}

\newcommand{\ev}[1]{\langle #1 \rangle}

\newcommand{\csw}{c_{\rm{sw}}}

\newcommand{\gtwo}{I\kern-.1em I\,}

\newcommand{\be}{\begin{equation}}
\newcommand{\ee}{\end{equation}}
\newcommand{\bea}{\begin{eqnarray}} 
\newcommand{\eea}{\end{eqnarray}}

\newcommand{\bmp}{\noindent\begin{minipage}{16cm}}
\newcommand{\emp}{\end{minipage}\vskip 7mm} 
\def\lsim{\mathrel{\raise.3ex\hbox{$<$\kern-.75em\lower1ex\hbox{$\sim$}}}}
\def\gsim{\mathrel{\raise.3ex\hbox{$>$\kern-.75em\lower1ex\hbox{$\sim$}}}}

\begin{document}
\begin{flushleft}{HIP-2012-02/TH}\end{flushleft}

\title{Perturbative Improvement of the Schr\"odinger Functional for Lattice Strong Dynamics}

\author{Tuomas Karavirta}
\email{tuomas.karavirta@jyu.fi} 
\affiliation{Department of Physics, University of Jyv\"askyl\"a, P.O.Box 35, FIN-40014 Jyv\"askyl\"a, Finland \\
and Helsinki Institute of Physics, P.O.Box 64, FIN-00014 University of Helsinki, Finland}
\author{Kari Rummukainen}
\email{kari.rummukainen@helsinki.fi}
\affiliation{Department of Physics and Helsinki Institute of Physics, 
P.O.Box 64, FIN-00014 University of Helsinki, Finland}
\author{Kimmo Tuominen}
\email{kimmo.i.tuominen@jyu.fi}
\affiliation{Department of Physics, University of Jyv\"askyl\"a, P.O.Box 35, FIN-40014 Jyv\"askyl\"a, Finland \\
and Helsinki Institute of Physics, P.O.Box 64, FIN-00014 University of Helsinki, Finland}

\begin{abstract}Lattice simulations on SU(2) and SU(3) gauge theories with matter fields in the fundamental, adjoint and two index symmetric representations are needed to determine if these theories are near or within the conformal window as required for their applications in beyond standard model phenomenology. Simulations with Wilson fermion action are subject to artifacts linear in the lattice spacing $a$, and must be improved. We provide the necessary coefficients for perturbative improvement of the boundary terms when using Schr\"odinger functional boundary conditions and furthermore show that correctly implemented ${\mathcal{O}}(a)$ improved actions are necessary for reliable results. 
\end{abstract}

\maketitle


There has recently been interest in studying quantum field theories
with a nontrivial infrared fixed point (IRFP), both in continuum and on
the lattice. Under the renormalization group evolution, the coupling 
of these theories shows asymptotic
freedom at small distances, analogously to QCD, but flows to a fixed
point at large distances where the theory hence looks conformal. Such theories have
applications in beyond Standard Model model building. These include
unparticles, i.e. an infrared conformal sector coupled weakly to the
Standard Model \cite{Georgi:2007ek}, and
(extended) technicolor scenarios, that explain the masses of the
Standard Model gauge bosons and fermions via strong coupling gauge
theory dynamics \cite{TC,Hill:2002ap,Sannino:2008ha}.
In addition to direct applications to particle phenomenology, the phase diagrams of 
gauge theories, as a function of the number of colours, $N$, flavours $N_f$ and fermion 
representations, are interesting from the purely theoretical viewpoint of understanding the 
nonperturbative gauge theory dynamics from first principles. 

Several methods to estimate the vacuum phase diagram of a gauge theory exist. 
A traditional method is the ladder approximation to the Schwinger-Dyson equation for the fermion 
propagator yielding an estimate for the onset of chiral symmetry breaking and signaling the departure 
from conformal to confining phase \cite{Appelquist:1998rb,Sannino:2004qp}. 
However, the only truly first principle method is 
constituted by lattice simulations. Several initial studies have appeared in literature: for example 
SU(2) with fundamental representation fermions \cite{Bursa:2010xr,Karavirta:2011zg}, SU(2) with adjoint 
fermions \cite{Catterall:2007yx,Hietanen:2008mr,DelDebbio:2008zf,
Catterall:2008qk,Hietanen:2009az,DeGrand:2011qd} and SU(3) with fermions in the fundamental  
\cite{Appelquist:2007hu,Fodor:2009wk,Deuzeman:2008sc} or in the two-index symmetric 
\cite{Shamir:2008pb}, i.e. the sextet, representation.

The studies with Wilson fermions are subject to lattice artifacts proportional to the lattice spacing $a$. 
In the context of high precision results for QCD like theories, i.e. SU(N) gauge theory with modest 
number of flavors, a program to cancel these lattice artifacts has been devised 
\cite{Sheikholeslami:1985ij,Luscher:1992an}. The basic idea in this approach is to introduce local 
counterterms whose coefficients will be fine tuned to cancel all ${\cal O}(a)$ contributions. 
In this Letter we provide the values of the perturbative coefficients required in the analyses of SU(2) 
and SU(3) gauge theories with fermions in the fundamental, adjoint and two-index symmetric 
representations. In addition to these results, we highlight their significance: The improved actions 
neglecting some of the counterterms will not work and, moreover, the boundary conditions for the 
background field when using Schr\"odinger functional, must be chosen carefully. This point has been emphasised also in \cite{Sint:2011gv}.

The basic Wilson lattice action is $S_{0}=S_{\textrm G} + S_{\textrm F}$, where
$S_{\textrm G}$ denotes the standard Wilson plaquette action and
$S_{\textrm F}$
%
is the usual Wilson fermion action for $N_f$ (mass degenerate) Dirac fermions in the 
fundamental or 
higher representation of the gauge group.
%
On lattices with periodic boundaries, the $\mathcal{O}(a)$ discretization errors in this action can be removed by introducing the 
Sheikholeslami-Wohlert -term \cite{Sheikholeslami:1985ij}  
\bea
S_{\rm{impr}} &=& S_0+a^5 c_{\textrm{sw}}\sum_x \bar\psi(x)\frac{i}{4}\sigma_{\mu\nu} F_{\mu\nu}(x)\psi(x),
 \label{swterm}
\eea
and tuning the coefficient $c_{\textrm{sw}}$ so that the ${\mathcal{O}}(a)$ effects in the on-shell 
quantities cancel. 
Here $\sigma_{\mu\nu}=i[\gamma_\mu,\gamma_\nu]/2$ and $F_{\mu\nu}(x)$ is the symmetrized lattice 
field strength tensor. 
 
The Schr\"odinger functional scheme is often used to measure the
evolution of the coupling constant.  In this scheme new contributions to linear lattice artifacts arise due to 
$\mathcal{O}(a)$ errors arising from the fixed spatially constant boundary  conditions at times $t=0$ and $t=L$.  
Here $L^4$ is the volume of the lattice, and the spatial link variables at the 
$t=0$ and $t=L$ boundaries are fixed to
\begin{equation} 
  \label{eq:abelian1}
  U_k(x)|_{(x_0=0)}=\exp(aC_k),\quad U_k(x)|_{(x_0=L)}=\exp(aC'_k),
\end{equation}
where $a$ is the lattice spacing and
\begin{equation} 
  C_k=\frac{i}{L}{\rm{diag}}(\phi_1,\dots,\phi_{N_c}),
  \quad C'_k=\frac{i}{L}{\rm{diag}}(\phi'_1,\dots,\phi'_{N_c}),
  \label{boundarymatrix}
\end{equation}   
with the constraint 
$\sum_{i=1}^{N_c} \phi_i = \sum_{i=1}^{N_c} \phi'_i = 0$.
For SU(2) and SU(3) the boundaries can be parametrized as
\cite{Luscher:1992zx, Luscher:1993gh}
\begin{equation}
  (\phi_1,\phi_2) = (\eta,-\eta), ~~
  (\phi'_1,\phi'_2) = (\rho-\eta,-\rho+\eta) ~~\mbox{SU(2)}
  \label{eq:CSU2}
\end{equation}
\begin{equation}
  \begin{array}{rcl}
   (\phi_1,\phi_2,\phi_3) &=& (\eta-\rho,-\eta/2,-\eta/2+\rho) \\
  (\phi'_1,\phi'_2,\phi'_3) &=& (-\eta-3\rho,\eta/2+\rho,\eta+2\rho) 
  \end{array}
  ~~~ \mbox{SU(3)}.
  \label{eq:CSU3}
\end{equation}
The conventional choice of parameters is $(\rho,\eta) = (\pi,\pi/4)$ for SU(2) and $(\rho,\eta)=(\pi/3,0)$ for SU(3). 
The spatial boundary conditions are taken to be periodic.
The fermion fields are set to vanish at the $t=0$ and $t=L$ boundaries and have twisted 
periodic boundary conditions in spatial directions: $\psi(x+L\hat e_i) = \exp(i\pi/5)\psi(x)$.  This improves the condition number of the fermion 
matrix \cite{Sint:1995ch}.
The values of $\phi_i$ in (\ref{boundarymatrix}) are restricted within
the so-called fundamental domain in order to guarantee a unique least action
solution for the background field \cite{Luscher:1992an,Luscher:1993gh}.

At the classical level the nontrivial boundary conditions generate a spatially constant chromoelectric 
field and the derivative of the action with respect to $\eta$ can be easily calculated;
it is proportional to the inverse of the bare coupling $g_0^2$.  At the full quantum
level the coupling can now be defined by  \cite{Luscher:1993gh}
\begin{equation}
  \frac{g_0^2}{g^2} = \ev{ \frac{\partial S}{\partial \eta} } {\bigg /}
  \frac{\partial S^{\textrm{cl.}}}{\partial \eta} 
  \label{sfcoupling}
\end{equation}
With this preliminary definition, the running of the coupling is quantified using the step scaling function:
\bea
\Sigma(u,s,L/a) &=& g^2(g_0,sL/a)\vert_{g^2(g_0,L/a)=u} \nonumber \\
&=& u+(\Sigma_{1,0}+\Sigma_{1,1} N_f)u^2.
\eea
The second line gives the formula in perturbation theory to one loop order, and 
the fermion contribution is denoted by $\Sigma_{1,1}$ while $\Sigma_{1,0}$ stands for the pure gauge one loop contribution. To evaluate these 
perturbative contributions we use the methods in 
\cite{Sint:1995ch,Sommer:1997jg}, and choose $s=2$. The continuum limit of $\Sigma_{1,1}$ ($\Sigma_{1,0}$) is given by the fermionic (bosonic) contribution to the one loop coefficient $b_{0}=\beta_0/(16\pi^2)$ of the beta function, where  $\beta_0=11/3 N_c-4/3 T(R) N_f$. In other words, 
\be
\delta_i=\lim_{L/a\rightarrow 0}\Sigma_{1,i}/(2 b_{0,i}\ln{2})=1,\,\, i=0,1
\label{deltadef}
\ee
where $b_{0,0}=11N_c/(48\pi^2)$ and  $b_{0,1}=N_fT(R)/(12\pi^2)$ corresponding, respectively, to bosonic and fermionic contribution in the beta function.

The $\mathcal{O}(a)$ errors originating from the nontrivial boundary field can be removed by introducing new 
terms to the action and fine tuning the corresponding coefficients so that the $\mathcal{O}(a)$ 
contributions cancel. Complete analysis of the necessary counterterms has been presented in 
\cite{Luscher:1996sc}. In the case of an electric background field and after setting the fermion 
fields to zero on the boundary, the required counterterms are of the form 
\bea
\delta S_{G,b}&=&-\frac{1}{g_0 ^2}(c_t-1)\sum_{p_t}{\rm{Tr}}[1-U(p_t)],\\
\delta S_{F,b}&=&a^4 (\tilde{c}_t-1)\sum_{\vec{x}}[\hat{O}_t(\vec{x})-\hat{O}'_t(\vec{x})].
\eea
Here $U(p_t)$ denotes plaquettes which touch the $t=0$ or $t=L$ boundaries.
The explicit from of the operators $\hat{O}_t$ and $\hat{O}^\prime_t$  is not needed in the following; for details, see  \cite{Luscher:1996sc}.
By tuning the coefficients $c_{\textrm{sw}}$, $ c_t$, $\tilde{c}_t$ to their proper values we can remove all  
$\mathcal{O}(a)$ errors from our action. Perturbatively,  the coefficient $\csw=1+{\cal{O}}(g_0^2)$, and it can be 
determined non-perturbatively in full lattice simulations. In the following we will concentrate on the one loop order
perturbative analysis of $c_t$ and $\tilde{c}_t$, and use $\csw=1$. 

The boundary coefficient $c_t$ has a perturbative expansion of the form
$
c_t=1+c_t ^{(1)} g_0 ^2+\mathcal{O}(g_0 ^4),
$
and similarly for $\tilde{c}_t$. To one loop order the coefficient $\tilde{c}_t$ can be 
extracted from the result of \cite{Luscher:1996vw} as
$\tilde{c}_t^{(1)}=-0.0135(1) C_F$ for fundamental fermions,
and this generalizes to other fermion representations simply by replacing the fundamental representation 
Casimir operator $C_F$ with Casimir operator $C_R$ of the representation $R$ under consideration. 
The results for different gauge groups and fermion representations are shown in table~\ref{table:pert_impro}.

The coefficient $c^{(1)}_t$ can be split into gauge and fermionic parts 
$c_t^{(1)}=c_t^{(1,0)}+c_t^{(1,1)}N_f$.
The contribution $c_t^{(1,0)}$ is entirely due to gauge fields and has been evaluated 
in \cite{Luscher:1992an} for SU(2) and in \cite{Luscher:1993gh} for SU(3). The fermionic contribution 
$c_t^{(1,1)}$ has been evaluated for fundamental fermions in \cite{Sint:1995ch} both for SU(2) and 
SU(3). We have extended these computations for SU(2) and SU(3) gauge theory with higher representation 
fermions \cite{Karavirta:2011mv}. The results for the nonzero improvement coefficients are tabulated in 
table \ref{table:pert_impro}. For the details of the numerical method used to determine 
coefficient $c_t^{(1,1)}$, we refer to the original literature where the method was developed for 
fundamental representation fermions \cite{Luscher:1992an,Sint:1995ch}. 
Our results are numerically consistent with the generic formula
$
c_t^{(1,1)} \approx 0.019141(2T(R)),
$
where $T(R)$ is the normalization of the representation $R$, defined as 
${\rm{Tr}}(T^a_R T^b_R)=T(R)\delta^{ab}$. 
\begin{table}[h!bt]
\center
\begin{tabular}{|c|c|c|c|c|}
\hline
$N_c$ & rep. & $c_t^{(1,0)}$ & $c_t^{(1,1)}$ &  $\tilde{c}_t^{(1)}$ \\
\hline
2 & ${\bf{2}}$ & $-0.0543(5)$ & $0.0192(2)$ & $-0.0101(3)$ \\
2 & ${\bf{3}}$ & $-0.0543(5)$ & $0.0766(2)$ & $-0.0270(2)$ \\
3 & ${\bf{3}}$ & $-0.08900(5)$ & $0.0192(4)$ & $-0.0180(1)$ \\
3 & ${\bf{8}}$ & $-0.08900(5)$ & $0.1148(3)$ & $-0.0405(3)$\\
3 & ${\bf{6}}$ & $-0.08900(5)$ & $0.09571(2)$ & $-0.0450(3)$\\
\hline
\end{tabular}
\caption{The nonzero improvement coefficients for Schr\"odinger functional boundary conditions with electric background field for various gauge groups and fermion representations.}
\label{table:pert_impro}
\end{table}
The numerical results collected in Table \ref{table:pert_impro} should be useful for studies of Wilson fermions in higher representations, and we highlight two issues: First, improvement should be carried out in order to obtain reasonably accurate results on lattices of reasonable size. Second, the improvement program has to be carried out consistently and correctly. 

\begin{figure}[htb]
\includegraphics[width=0.53\textwidth]{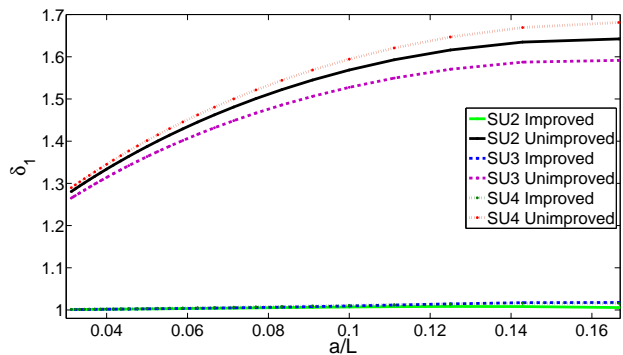}
\caption{Comparison between unimproved and improved results for the fermionic contribution to the perturbative step scaling function for SU(2), SU(3) and SU(4) gauge theories with fundamental representation matter fields.}
\label{fundamental}
\end{figure}

To illustrate the need for improvement, in Fig. \ref{fundamental} we show  $\delta_1$ defined in \eqref{deltadef}, to  compare the unimproved and improved results for SU(N) gauge theory with fundamental fermions for 
$N=2$,  $3$ and $4$. From the figure we see that improvement is essential for taming the ${\cal{O}}(a)$ artefacts. 

\begin{figure}[htb]
\includegraphics[width=0.53\textwidth]{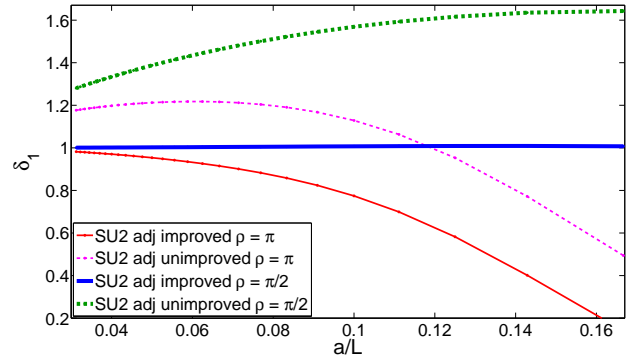}
\caption{Comparison between two different boundary fields in SU(2) gauge theory with adjoint fermions. The thin solid and dashed curves correspond, respectively, to improved and unimproved results with $\rho=\pi$. The thick solid (dashed) curves correspond to
improved (unimproved) results for $\rho=\pi/2$.
}
\label{phivaried}
\end{figure}

Next, in addition to using the coefficients in Table \ref{table:pert_impro} the precise values of the boundary fields must be 
chosen carefully to guarantee rapid convergence of the results. In Fig. \ref{phivaried} we show the results for $
\Sigma_{1,1}$ for SU(2) gauge theory and adjoint fermions. The results are normalized to the continuum values 
While the improvement is necessary to get rid of ${\cal{O}}(a)$ corrections, clearly,
in order to guarantee that 
the ${\cal{O}}(a^2)$ corrections remain small the boundary values must be optimized.  The optimal choice is $\rho=\pi/2$ and $0<\eta<\pi/2$ excluding the value $\eta=\pi/4$.
This can be understood as follows: If the boundary field is generically a diagonal matrix with eigenvalues $\phi$ and $-\phi$, then the adjoint fermions see these eigenvalues as $2\phi$, $-2\phi$ and 0.
In other words, since the background appears twice as large for adjoint fermions as for fundamental fermions, it seems 
plausible that the problems could be alleviated by halving the background field. We note that the numerical results imply that
with the above choice of parametrizing the boundary fields, the parameter $\rho$ plays the main role; i.e. there is a single value 
of $\rho$ (e.g. $\pi/2$ for the adjoint representation of SU(2)) for which the convergence is optimal and $\eta$ can be chosen from a wide interval without significantly altering the result. 

\begin{figure}
\includegraphics[width=0.54\textwidth]{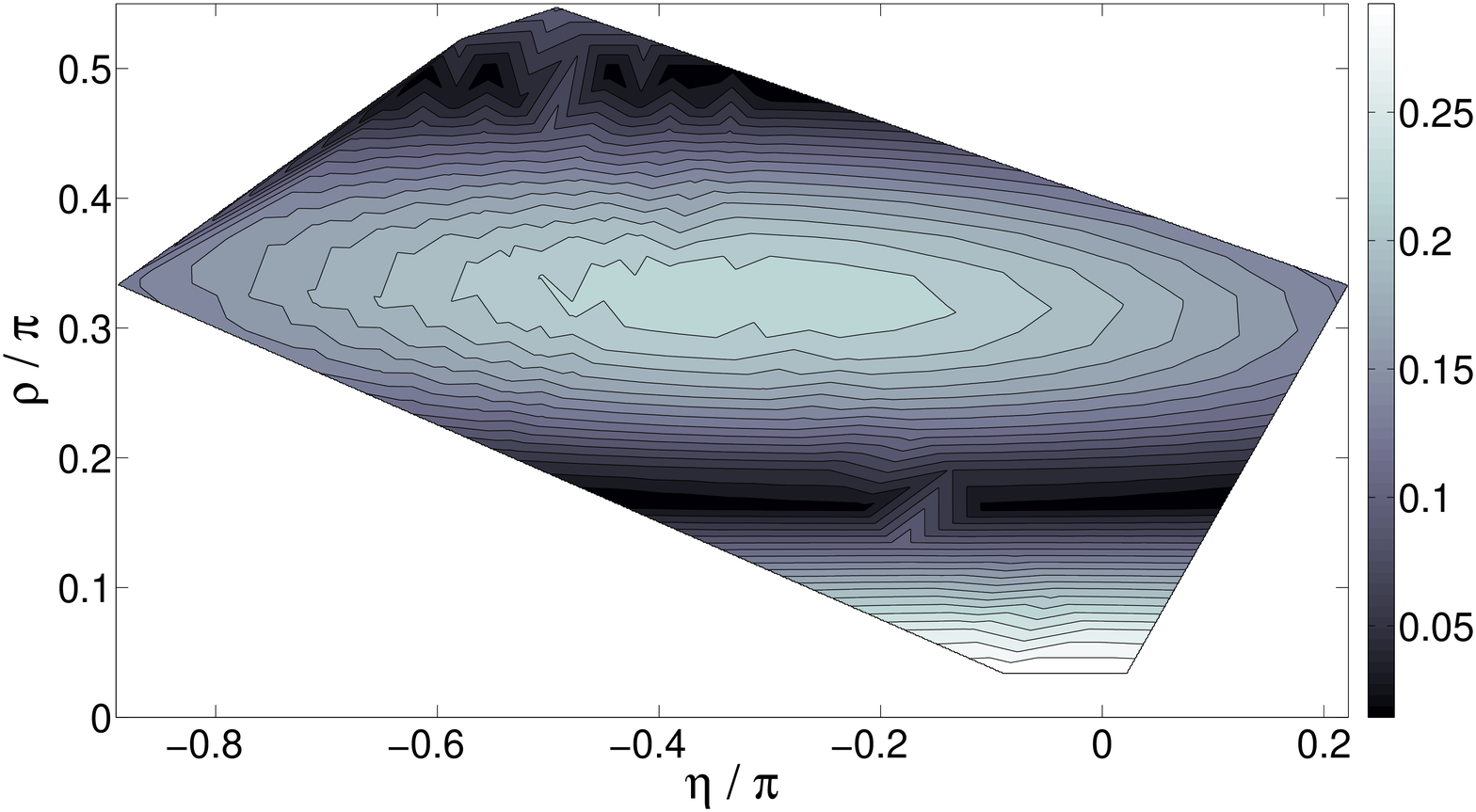}\\
\includegraphics[width=0.54\textwidth]{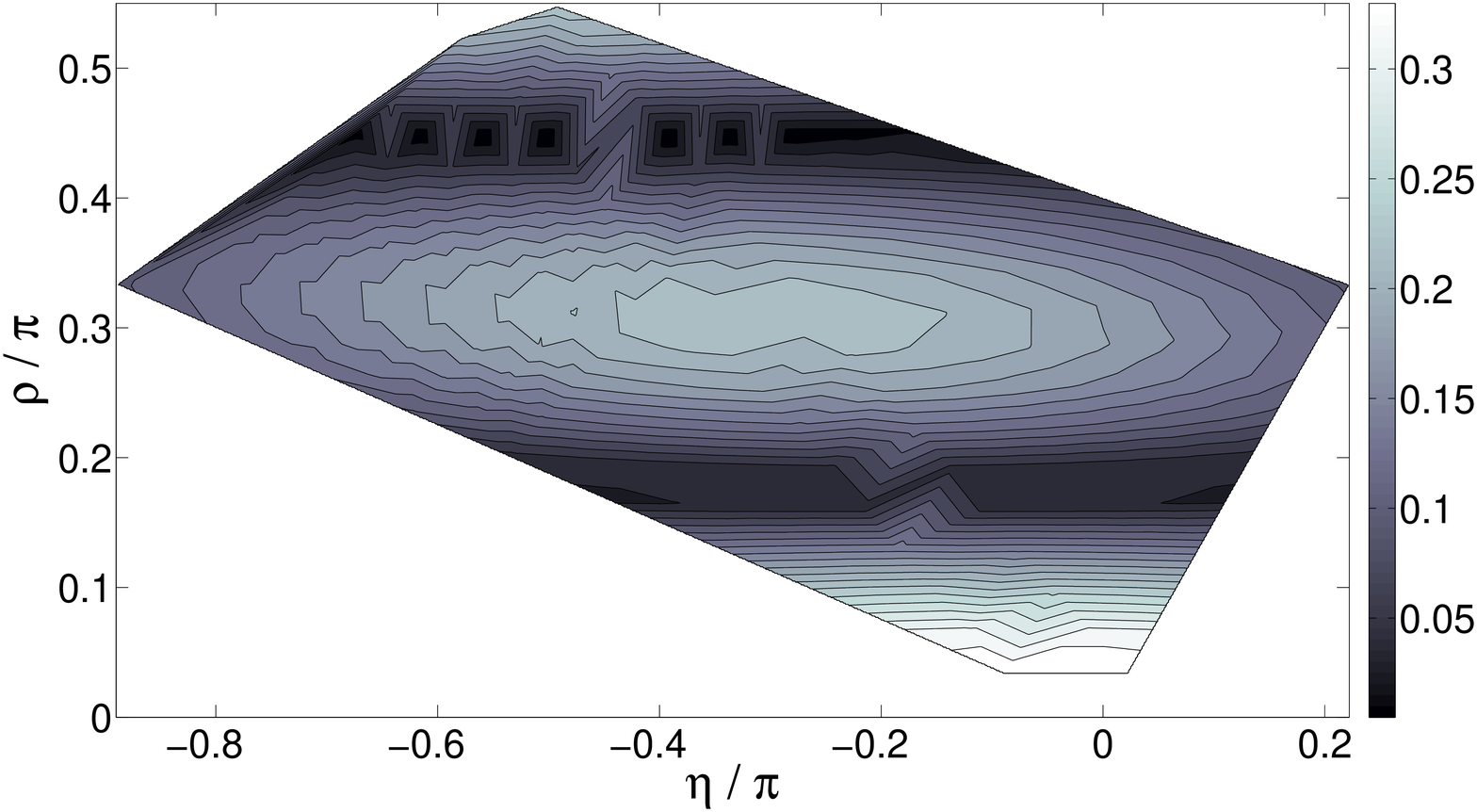}\\
\caption{Convergence of $\vert\delta_1-1\vert$ as a function of $\rho$ and $\eta$ for adjoint (upper panel) and sextet (lower panel) 
fermions in SU(3) gauge theory.}
\label{highrepsu3}
\end{figure}

The analysis is straightforward also for SU(3) with higher representation fermions. 
In Fig. \ref{highrepsu3} we show $\vert\delta_1-1\vert$ 
contours in the $(\eta,\rho)$-plane. The conventional choice for the fundamental representation fermions is $(\rho,\eta)=(\pi/3,0)$.
As in the case of SU(2), when translating the boundary matrix to the adjoint or sextet representation, one sees that the background 
field is effectively doubled. This again suggests trying to narrow the values of the angular parameters by a factor of two. This is indeed
what we observe from Fig. \ref{highrepsu3}: For the adjoint representation, shown in the upper panel, the best convergence of 
$\delta_1$ is obtained for $\rho=\pi/6$ and $-4\pi/9<\eta<\pi/9$ excluding the value $\eta=-\pi/6$. 
Similarly, as shown in the lower panel of the figure, for the sextet representation we find that the convergence is optimal for 
$\rho=67\pi/150\approx 0.45 \pi$ and $-166\pi/225<-7\pi/45$ excluding $\eta=-67\pi/150$. As for SU(2), we find that the 
convergence is dominantly controlled by the value of the $\rho$ while $\eta$ can be chosen from a wide interval. 
For the figure we used $L/a=10$, but the results remain quantitatively similar for 
other values of $L/a$ as well; the dependence on $L/a$ is similar to the SU(2) case shown in Fig. \ref{phivaried}.

\begin{figure}
\includegraphics[width=0.53\textwidth]{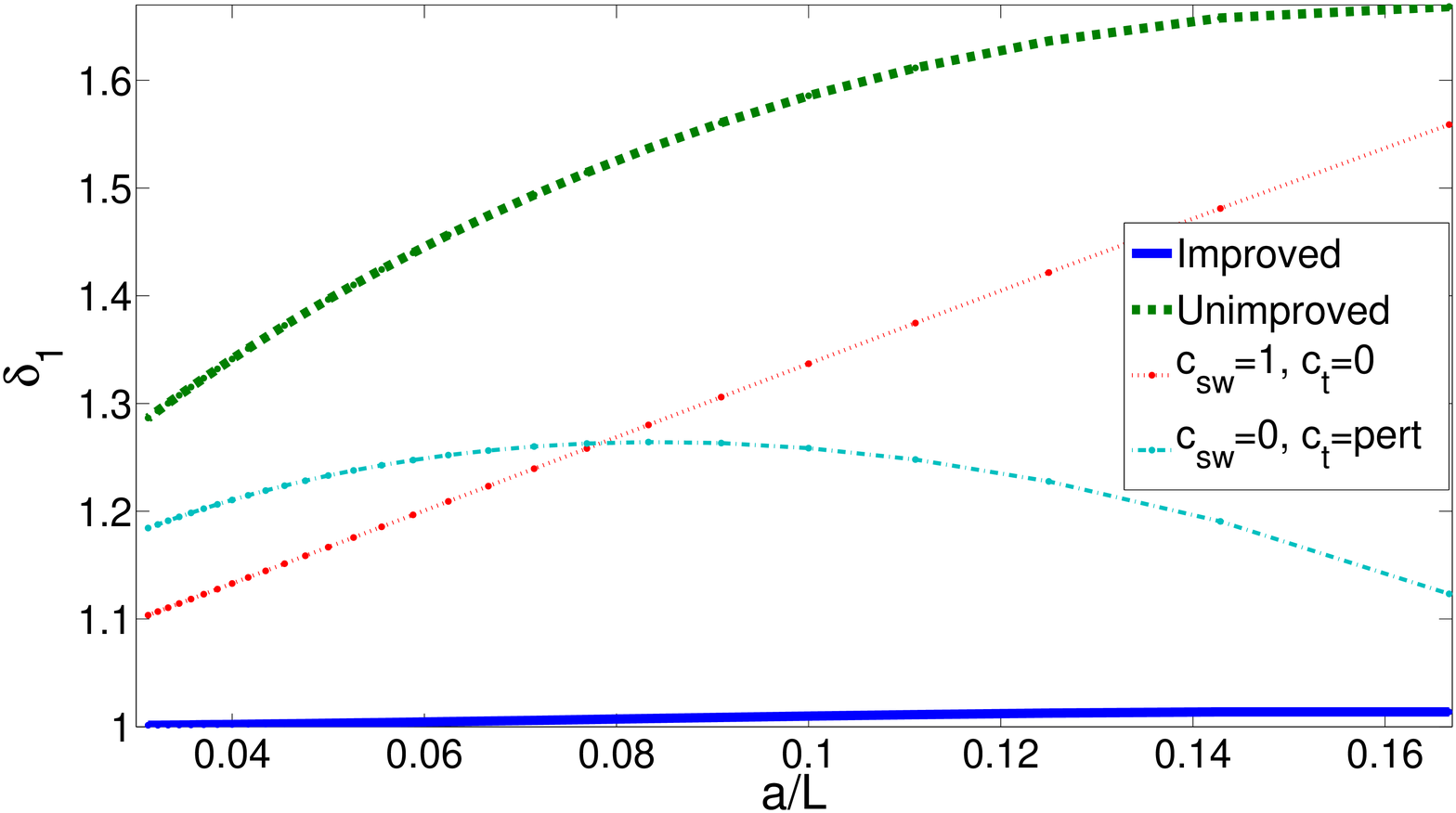}\\
\includegraphics[width=0.53\textwidth]{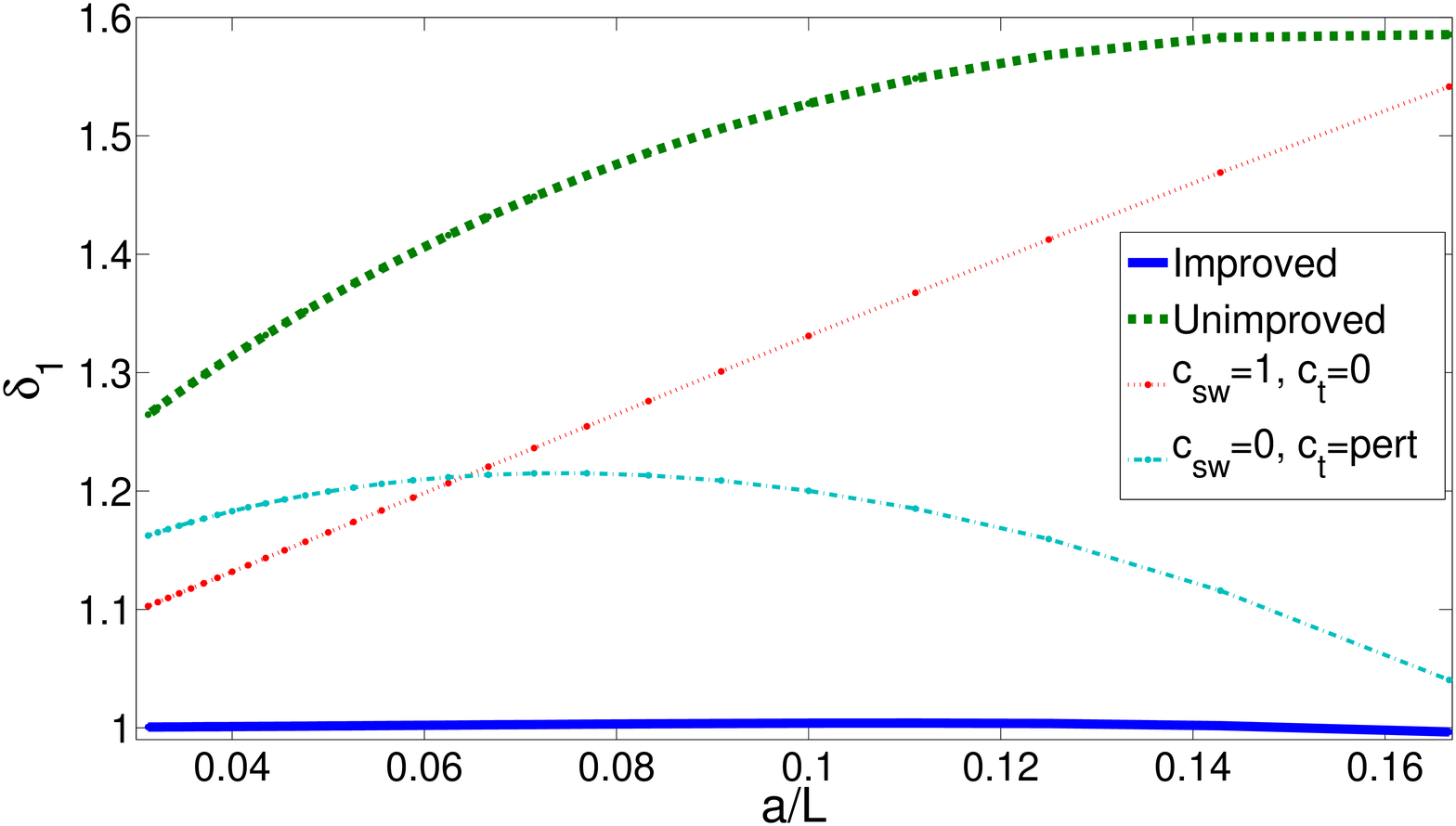}\\
\caption{The upper panel shows the result for SU(3) with adjoint fermions and the lower one for SU(3) with sextet fermions. The thick solid and dashed curves correspond, respectively, to improved and unimproved results. The thin dotted (dash-dotted) curves correspond to the results obtained by setting $\csw=1\, (0)$ and $c_t=0$ (perturbative value).}
\label{highreps}
\end{figure}

Finally, the improvement coefficients must be included consistently. In Fig \ref{highreps} we show comparison between the improved, unimproved results as well as between the results neglecting some of the improvement coefficients. In particular we see that even if the Sheikholeslami-Wohlert coefficient $\csw=1$ is implemented, but the boundary improvement is neglected, the results obtained under Schr\"odinger functional calculations like in \cite{DeGrand:2011qd} are possibly still far from the proper continuum limit.  Also the optimal values for the boundary fields used for the improved results in Fig. \ref{highreps} differ from the values used for the fundamental representation as we discussed earlier. We have used here the values of $\rho$ and $\eta$ which give optimal convergence for $\delta_1$ as discussed in previous paragraphs.


In this Letter we have described developments in analyses of gauge theories with higher representation matter fields on the lattice. Generally, we have demonstrated the sensitivity of the results on the correct values of the improvement coefficients and boundary conditions when using Wilson fermions. We have provided perturbative values of the counterterm coefficients required for the analysis and emphasized the choice of the boundary conditions. Our results suggest that in studies of higher representations there is need to
optimize the value of the background field carefully. This need is driven by the fermionic contribution; we have explicitly checked that 
the effect on the pure gauge contribution is negligible, on the level of few percents, for the cases we have considered. 

\begin{acknowledgments}
K.R. acknowledges support from the Academy of Finland grant 1134018.
\end{acknowledgments}


\end{document}